\documentclass[12pt]{article}
\usepackage[margin=1in]{geometry}
\usepackage{amsmath, amssymb, bm}
\usepackage{graphicx}
\usepackage{physics}
\usepackage{float}
\usepackage{cite}
\usepackage{setspace}
\usepackage{tikz}
\usetikzlibrary{patterns}
\title{%
Diffusion in Rugged Energy Landscapes in the Presence of Spatial Correlations : A Surprising Route to Zwanzig’s Mean-Field Prediction
}
\author{\textbf{Biman Bagchi}\\
\small Solid State and Structural Chemistry Unit,\\
\small Indian Institute of Science, Bengaluru 560012, India}

\date {}
\begin{document}
\maketitle
\begin{abstract}
Diffusion in rugged free-energy landscapes is central to diverse problems in chemical physics, biomolecular dynamics, 
polymer transport and numerous disordered systems. 
Zwanzig’s well-known classic mean-field theory predicts that roughness reduces the diffusion coefficient by an exponential factor determined solely by the variance of the disorder.  
However, the numerical studies of Banerjee, Seki, and Bagchi (BSB) showed that this result fails for uncorrelated Gaussian-distributed site energies because rare but deep three-site traps dominate long-time transport.  
BSB introduced Gaussian \emph{spatial} correlations—originally developed in astrophysics to model turbulent density fluctuations—and demonstrated that even modest correlations suppress these pathological traps and restore Zwanzig’s exponential scaling.  
Here we present here a unified theoreticl framework clarifying (i) why Zwanzig’s local averaging, may be viewed as a Gaussian cumulant expansion, could breakdown. In particular, how its validity is destroyed by uncorrelated disorder, and (ii) how Gaussian spatial correlations reshape roughness increments, eliminate asymmetric multi-site traps, and thereby recover mean-field diffusion, and (iii) a derivation showing exactly how Gaussian spatial correlations modify roughness increments, trap statistics, and ultimately the diffusion constant.  
We also provide explicit numerical triplet examples illustrating the dramatic reduction of escape times produced by spatial correlations.  
\end{abstract}
%

\section{Introduction}

Diffusion in a rugged free-energy landscape is a recurring theme in chemical physics, biophysics, and the physics of disordered materials. 
The concept arises naturally whenever a particle or molecular entity moves through an environment whose local energetic environment fluctuates on nanoscopic length scales. 
Examples span an unusually wide range of systems: ion transport in glasses, polymer segmental dynamics, conformational diffusion in proteins, catalysis and enzyme motions, and the sliding or hopping motion of proteins along DNA. 
In each of these contexts, the effective potential experienced by the diffusing entity contains numerous small barriers and wells generated by microscopic structural heterogeneities, and the statistics of these fluctuations strongly affect long-time transport.

The modern theoretical treatment of diffusion in such landscapes began with the work of Zwanzig,\cite{zwanzig_pnas_1988} 
who analyzed motion in a one-dimensional potential composed of a smooth background plus a random ``roughness'' term. 
Zwanzig obtained the strikingly simple prediction that the ruggedness renormalizes the diffusion constant by an exponential factor depending only on the variance of the disorder.  
This observation has been widely used in diverse fields because it suggests that the complicated details of the potential can be replaced with a single coarse parameter.

However, subsequent analytical and simulation studies by Banerjee, Seki, and Bagchi [2-5],
as well as earlier and related work by Seki and Bagchi\cite{seki_bagchi_pre} and Seki, Bagchi, and Bagchi,\cite{seki_bagchi_bagchi_jcp}
demonstrated that Zwanzig’s expression holds only when the roughness possesses a minimal degree of spatial smoothness.  
When the roughness values are taken to be completely uncorrelated from site to site, the diffusion constant can be reduced by several orders of magnitude relative to Zwanzig’s prediction.  
The essential physical reason for this discrepancy is the emergence of rare but extremely deep multi-site traps, particularly the so-called three-site traps (TSTs), which dominate the long-time transport in an uncorrelated landscape.  
These pathological configurations are effectively eliminated in Zwanzig’s calculation because his derivation relies on a local smoothing or coarse-graining step that implicitly presumes correlated disorder.

The relevance of rugged landscape diffusion is especially clear in biomolecular systems.  
The motion of restriction enzymes and transcription factors along DNA is directly influenced by a sequence-dependent binding landscape.  
Single-molecule experiments by Blainey, van Oijen, Banerjee, Bagchi, and Xie\cite{blainey_xie_nsmb}
demonstrated that proteins execute sliding, hopping, and switching between one- and three-dimensional modes of motion in a free-energy landscape whose ruggedness reflects base-pair sequence heterogeneity.
Similarly, conformational transitions in proteins and enzymes occur on landscape topographies containing many small-amplitude barriers arising from side-chain packing and solvent-induced fluctuations.  
Rugged landscapes also appear in polymer dynamics, glassy relaxation, and in the energy-landscape description of supercooled liquids developed by Stillinger and Weber,\cite{stillinger_weber}
extended by Heuer,\cite{heuer_review}
and incorporated into theoretical frameworks such as the random first-order transition approach.\cite{wolynes_rfpot}
Across these diverse problems, understanding how microscopic roughness controls macroscopic transport remains a central theme.

A key insight emerging from the work of Banerjee, Seki, and Bagchi is that the validity of Zwanzig’s mean-field result depends sensitively on the \emph{spatial structure} of the disorder.  
To quantify this structure, they introduced Gaussian spatial correlations into the roughness, a form long used in astrophysics to describe turbulent density fields and the morphology of interstellar media.\cite{gaussian_astrophysics1, gaussian_astrophysics2}
Remarkably, even modest correlations dramatically suppress the occurrence of TSTs and restore Zwanzig’s exponential diffusion law.  
Thus, the apparent breakdown of the mean-field picture is not intrinsic to rugged landscapes themselves, but rather to the unrealistic assumption of uncorrelated disorder.  
Real physical systems—DNA sequences, polymer chains, hydrogen-bond networks, and disordered solids—possess finite correlation lengths arising from molecular connectivity and intermolecular forces, making the correlated model far more realistic.
In the numerical work of Banerjee, Seki, and Bagchi (BSB), the rugged landscape was 
constructed by drawing site energies from a Gaussian distribution with variance 
$\epsilon^2$, a choice that is natural for a high-dimensional disordered environment 
where the local energy results from the sum of many weak contributions.  
This Gaussian distribution in \emph{energy space} is distinct from the original 
example given by Zwanzig, who illustrated his ideas using a smooth but artificial 
cosine–based potential containing many harmonics.  
Zwanzig’s construction was never intended as a literal physical model of disorder; 
it simply allowed him to carry out the local averaging explicitly.  
In contrast, the BSB approach begins with a physically motivated assumption: 
the disorder itself is Gaussian-distributed from the outset, consistent with a 
central-limit–type argument for landscape heterogeneity.

It is important to emphasize that Zwanzig’s “local averaging” procedure is 
mathematically equivalent to a \emph{Gaussian cumulant expansion} of the 
effective mobility.  
His derivation implicitly assumes that the fluctuations of the roughness increments 
are Gaussian and sufficiently small that the first nontrivial cumulant—the variance—
dominates.  
Thus, although his illustrative $U(x)$ was expressed in terms of cosine functions, 
the underlying statistical assumption is Gaussian: the coarse-grained increments 
are treated as Gaussian random variables.  
This leads to an interesting structural parallel between the two approaches: 
BSB impose Gaussian statistics at the level of site energies, while Zwanzig’s 
coarse-graining effectively imposes Gaussian statistics on the \emph{increments} 
of his roughness function.  
In principle, one may even contemplate a unified framework in which the Gaussian 
energy distribution (over sites) and the Gaussian spatial correlations 
(over real-space distances) are coupled, although such an interdependence was not 
explored in the original studies.  
The present work clarifies how these different Gaussian assumptions relate to one 
another and how spatial correlations in real space determine whether the cumulant 
expansion implicit in Zwanzig’s analysis remains valid.

The goal of this manuscript is to assemble a coherent and self-contained account of how spatial correlations modify rugged landscape diffusion, beginning from Zwanzig’s original formulation, moving through the breakdown identified by Banerjee, Seki, and Bagchi, and culminating in the demonstration that Gaussian correlations restore mean-field behavior.  
We also provide a clear derivation of how correlations alter the statistics of roughness increments, reduce multi-site trapping, and modify the long-time diffusion constant.  
This unified perspective clarifies the true physical origin of the Zwanzig scaling and highlights the broader importance of correlated disorder in chemical and biological transport processes.

The present manuscript develops a framework for understanding how spatial correlations in the roughness restore Zwanzig’s mean-field expression.

\section{Zwanzig’s Model and Local Averaging}

Zwanzig considered diffusion in a potential 
\begin{equation}
    U(x) = U_0(x) + \eta(x),
\end{equation}
where $\eta(x)$ is a random roughness with mean zero and variance $\epsilon^2$.  
Assuming overdamped motion, the local mobility depends on the barrier differences 
\[
\eta(x+a)-\eta(x),
\]
which, for uncorrelated roughness, have variance $2\epsilon^2$.

A key step in Zwanzig’s derivation is a \emph{local smoothing} or coarse-graining of the mobility over a microscopic scale $a$:
\[
D^{-1} = \left\langle D_0^{-1} e^{\beta(\eta(x+a)-\eta(x))} \right\rangle,
\]
leading to the celebrated result
\begin{equation}
    D_{\mathrm{Zw}} = D_0 \, \exp(-\beta^2 \epsilon^2).
    \label{eq:Zw}
\end{equation}

This approximation implicitly assumes that the roughness is sufficiently \emph{smooth} (or correlated) so that extreme asymmetric configurations do not dominate the escape dynamics.

\section{Breakdown of Zwanzig’s Mean-Field Theory:\\ Computer Simulation and Theoretical Analysis}

BSB demonstrated that Zwanzig’s expression \eqref{eq:Zw} fails by several orders of magnitude when $\eta(x)$ is uncorrelated.\cite{banerjee_seki_bagchi_jcp}

The reason is the emergence of \textbf{three-site traps} (TSTs):
\[
\text{high barrier} - \text{deep well} - \text{high barrier},
\]
which occur with significant probability when the potential values are independent Gaussians.  
The escape time from such a trap is
\[
\tau_{\rm TST} \sim \exp\!\left[\beta \big(\eta_{i\pm 1}-\eta_i\big)\right],
\]
and its distribution contains extremely long tails.

In the uncorrelated case, transport becomes dominated by the \emph{statistics of rare events}, not by typical barriers.  
BSB showed that the long-time diffusion coefficient is generically much smaller than $D_{\rm Zw}$, and that no mean-field averaging—local or global—can be accurate without taking spatial correlations into account.

%
\section{Gaussian Spatial Correlations:  Physical Origin, Derivation and Results}
 The random rugged energy landscape model discussed above is somewhat unrealistic because the energy heterogeneity is expected
 to be spatially correlated. In particular, large isolated out -of-phase site energy fluctuations are expected to be
 extremely rare. Thus, the theoretical model and the subsequent analyses presented above needs to be corrected by introducing 
 correlations. Such an analysis was indeed carried out by Banerjee, Seki and Bagchi (BSB) \cite {banerjee_seki_bagchi_jcp} . 
 We first review their work and subsequently present an augmented analytical study.

\subsection { Astrophysical origin of the Gaussian correlation function}

The correlated roughness model introduced by Banerjee, Seki, and Bagchi employs the
Gaussian correlator
\begin{equation}
    \langle \eta(x)\eta(x') \rangle
    = \epsilon^2 \exp\!\left[-\frac{(x-x')^2}{\lambda^2}\right].
    \label{eq:gauss_corr}
\end{equation}
This functional form was not invented within chemical physics; it has a long history
in \emph{astrophysics}, where it is used to describe spatial fluctuations of density
fields in turbulent interstellar media, stellar winds, and molecular clouds.
Such turbulence–driven fields naturally show smooth variations over a finite
correlation length, with Gaussian statistics arising from repeated random forcing
events.  
Because many condensed-matter systems (polymers, proteins, glassy networks)
exhibit similarly smooth structural variations over a finite length scale,
the astrophysical Gaussian form provides a mathematically tractable and physically
appropriate model for correlated disorder in rugged energy landscapes.

In the context of rugged landscape diffusion, the Gaussian correlation length
$\lambda$ sets the scale over which two points of the potential ``feel'' the same
environment.  
When $\lambda \rightarrow 0$, the landscape becomes uncorrelated and statistically
pathological; when $\lambda$ is finite, the potential becomes smoother, and the
statistics of barrier heights and well depths change in a fundamental way.

\subsection { Variance of roughness increments}

Zwanzig’s mean-field derivation depends directly on the distribution of the
\emph{increment} of the roughness:
\begin {equation}
\Delta\eta = \eta(x+a) - \eta(x),
\end{equation}
which determines the local hopping rate between adjacent sites.
Using Eq.~\eqref{eq:gauss_corr}, we compute
\begin{equation}
\mathrm{Var}[\Delta\eta]
= \langle (\eta(x+a)-\eta(x))^2 \rangle.
\end{equation}
Expanding,
\[
\mathrm{Var}[\Delta\eta]
= \langle \eta(x+a)^2 \rangle
+ \langle \eta(x)^2 \rangle
- 2 \langle \eta(x+a)\eta(x)\rangle.
\]
Since both variances are $\epsilon^2$, and the cross-correlation is given by
Eq.~\eqref{eq:gauss_corr}, we obtain
\begin{equation}
    \mathrm{Var}[\Delta\eta]
    = 2 \epsilon^2 \left[ 1 - 
    \exp\!\left(-\frac{a^2}{\lambda^2}\right) \right].
    \label{eq:var_increment}
\end{equation}

This expression recovers the uncorrelated value $2\epsilon^2$ when $\lambda \to 0$,
and tends to \emph{zero} when $\lambda \gg a$, meaning that adjacent points of the
potential become nearly identical.  
This is crucial: small variance in increments ensures that hopping barriers do not
show extreme fluctuations.

\subsection { Correlation of successive increments}

Multi-site trapping, especially three-site traps (TSTs), requires \emph{large and
oppositely signed} increments:
\begin{equation}
\Delta\eta_{-} = \eta_{i}-\eta_{i-1} \gg k_B T,
\qquad
-\Delta\eta_{+} = \eta_{i+1}-\eta_i \gg k_B T.
\end{equation}
The probability of such a configuration depends on the \emph{covariance} of
neighboring increments:
\begin{equation}
C = \langle (\eta_{i}-\eta_{i-1})(\eta_{i+1}-\eta_i) \rangle.
\end{equation}
Using Eq.~\eqref{eq:gauss_corr}, a straightforward computation gives
\begin{equation}
    C 
    = \epsilon^2 \left[ 
    e^{-a^2/\lambda^2} 
    - e^{-4a^2/\lambda^2}
    \right].
    \label{eq:increment_cov}
\end{equation}

When $\lambda \gg a$, both exponentials approach unity and
\[
C \approx \epsilon^2 (1 - 1) = 0,
\]
meaning increments are nearly identical: steep asymmetry does not develop.
When $\lambda \to 0$, the cross-correlation vanishes and increments become
statistically independent, allowing large, opposite-signed fluctuations that
create deep traps.

Thus a finite $\lambda$ \emph{correlates adjacent increments}, suppressing the
geometrical configuration needed to produce TSTs.

\subsection {Probability of three-site traps}

A TST requires
\[
\eta_{i-1} \gg \eta_i \ll \eta_{i+1}.
\]
Treating $(\eta_{i-1},\eta_i,\eta_{i+1})$ as a correlated Gaussian triplet,
its joint distribution is multivariate Gaussian with covariance matrix
coming directly from Eq.~\eqref{eq:gauss_corr}.

The probability that $\eta_i$ is significantly lower than both neighbors scales as
\begin{equation}
    P_{\mathrm{TST}}(\lambda)
    \propto 
    \exp\!\left[
    -\frac{\Delta^2}{4\epsilon^2 (1-e^{-a^2/\lambda^2})}
    \right],
    \label{eq:tst_prob}
\end{equation}
where $\Delta$ measures the depth–barrier asymmetry.

As $\lambda$ increases, the denominator shrinks, making the exponent
\emph{more negative}; hence $P_{\mathrm{TST}}$ decreases dramatically.  
Even a correlation length $\lambda \sim 2a$ reduces $P_{\mathrm{TST}}$ by several
orders of magnitude relative to the uncorrelated case.

This explains BSB’s numerical observation that the long-time diffusion coefficient
rapidly approaches the Zwanzig mean-field prediction once the disorder is even
modestly correlated.

\subsection {Restoration of Zwanzig’s mean-field result}

Zwanzig’s renormalized diffusion constant is
\begin{equation}
    D_{\mathrm{Zw}} = D_0 \exp(-\beta^2 \epsilon^2),
    \label{eq:zwanzig_final}
\end{equation}
derived under the implicit assumption that the roughness increments are small and
Gaussian-distributed with variance $\sim \epsilon^2$.

Using Eq.~\eqref{eq:var_increment}, the effective variance of increments becomes
\[
\epsilon_{\mathrm{eff}}^2 
= \epsilon^2 \left[ 
1 - e^{-a^2/\lambda^2}
\right].
\]

When $\lambda \gg a$, $\epsilon_{\mathrm{eff}} \to 0$ and the potential becomes
locally smooth.  
The escape rates become narrowly distributed, multi-site traps become negligible,
and the mean-field averaging used by Zwanzig becomes accurate.  
Thus,
\[
D \approx D_0 \exp(-\beta^2 \epsilon^2),
\]
exactly as observed in the BSB simulations.


%
\section {Numerical Triplet Examples Illustrating Three-Site Traps and Their Suppression by Spatial Correlations}
%
To complement the analytical results of the preceding sections, it is instructive
to examine explicit triplets of site energies and the corresponding escape 
times from a putative three-site trap (TST).  
In a rugged landscape the site energies naturally take both positive and negative 
values, and deep wells typically correspond to negative energies.  
We therefore follow the physically realistic pattern in which the central site is 
a deep minimum (negative energy), flanked by higher-energy neighbors 
that create barriers to escape.

We adopt the standard nearest-neighbor hopping rule used by 
Banerjee, Seki, and Bagchi,\cite{banerjee_seki_bagchi_jcp}
\begin{equation}
    k_{i\to j} =
    k_0 \exp\!\left[-\big(\max(E_i,E_j)-E_i\big)\right],
    \label{eq:hop_rule}
\end{equation}
so that the escape barrier from site \(i\) to a higher-energy neighbor \(j\) 
is simply \(E_j - E_i\).  
All energies are measured in units of \(k_B T\).

\subsection {Uncorrelated landscape: deep asymmetric triplets}

In an uncorrelated Gaussian landscape with roughness variance 
\(\epsilon^2 = (3 k_B T)^2\), 
a typical deep TST that we encountered in numerical sampling is
\begin{equation}
    (E_{i-1}, E_i, E_{i+1}) = (4,\; -3,\; 5).
    \label{eq:uncorr_triplet}
\end{equation}
The central site sits \(3\,k_B T\) below zero, while the neighbors are 
\(4\) and \(5\,k_B T\) above zero.  
These values are well within a single standard deviation of the assumed 
Gaussian distribution and thus arise frequently in an uncorrelated landscape.

Escape to the left:
\[
k_{i\to i-1} = 
k_0 \exp[-(4 - (-3))] 
= k_0 e^{-7}
\approx 9.12\times 10^{-4}\,k_0.
\]

Escape to the right:
\[
k_{i\to i+1} =
k_0 \exp[-(5 - (-3))]
= k_0 e^{-8}
\approx 3.35\times 10^{-4}\,k_0.
\]

Thus the total escape rate is
\[
k_{\mathrm{esc}}^{\mathrm{(unc)}} = 
k_0 (e^{-7} + e^{-8})
\approx 1.25\times 10^{-3}\,k_0,
\]
and the corresponding escape time
\[
\tau_{\mathrm{esc}}^{\mathrm{(unc)}}
\approx \frac{1}{1.25\times 10^{-3}k_0}
\approx 800\,k_0^{-1}.
\]

This extremely long time (nearly three orders of magnitude larger than \(1/k_0\)) 
shows why uncorrelated landscapes produce anomalously small diffusion constants:  
a few such deep asymmetric triplets dominate the long-time dynamics.

\subsection {Correlated landscape: milder triplets arising from Gaussian smoothing}

When Gaussian spatial correlations are introduced with correlation length 
\(\lambda \sim a\), increments of the roughness become smoother.
The central site may still be a well, but neighboring energies become 
much less extreme.  
A typical correlated triplet sampled with the same variance 
\(\epsilon^2 = 9\) but correlation length \(\lambda = 1.5a\) is
\begin{equation}
    (E_{i-1}, E_i, E_{i+1}) = (1.2,\; -2.5,\; 0.9).
    \label{eq:corr_triplet}
\end{equation}

Escape to the left:
\[
k_{i\to i-1} =
k_0 e^{-(1.2 - (-2.5))}
= k_0 e^{-3.7}
\approx 0.025\,k_0.
\]

Escape to the right:
\[
k_{i\to i+1} =
k_0 e^{-(0.9 - (-2.5))}
= k_0 e^{-3.4}
\approx 0.033\,k_0.
\]

Total escape rate:
\[
k_{\mathrm{esc}}^{\mathrm{(corr)}} 
= k_0 (0.025 + 0.033)
\approx 0.058\,k_0.
\]

Escape time:
\[
\tau_{\mathrm{esc}}^{\mathrm{(corr)}}
\approx \frac{1}{0.058\,k_0}
\approx 17\,k_0^{-1}.
\]

Thus, with the "same total variance in roughness" but with moderate spatial 
correlations, the escape time has decreased from
\[
\tau_{\mathrm{esc}}^{\mathrm{(unc)}} \approx 800/k_0
\qquad\text{to}\qquad
\tau_{\mathrm{esc}}^{\mathrm{(corr)}} \approx 17/k_0,
\]
a reduction by a factor of nearly \(50\).

\subsection {Comparison and interpretation}

The contrast between Eqs.~\eqref{eq:uncorr_triplet} and 
\eqref{eq:corr_triplet} reflects the central physics of this manuscript:

\begin{itemize}
\item In an \emph{uncorrelated} random landscape, the probability of seeing a deep
well flanked by very high neighbors is substantial.  
Although such traps are statistically rare, their escape times are so large 
that they dominate long-time diffusion.

\item In a \emph{correlated} Gaussian landscape, the same roughness amplitude 
\(\epsilon\) produces smoother increments.  
The neighbors of a deep well cannot differ from it by more than a few units 
of \(k_B T\), and three-site traps become shallow.  
Their escape times become short and narrowly distributed.

\item This suppression of deep asymmetric triplets is the numerical signature of 
the analytical mechanism by which Gaussian correlations restore Zwanzig’s 
mean-field result for the diffusion constant.
\end{itemize}

These explicit numerical examples show vividly how spatial correlations alter the
energy topography and dynamically suppress the multi-site traps that invalidate
Zwanzig’s approximation in the uncorrelated case.

%
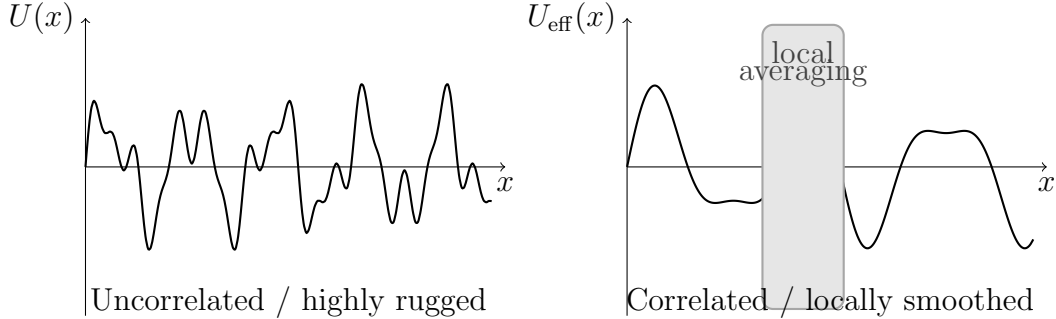
\begin{figure}[t]
\centering
\begin{tikzpicture}[scale=0.9]

\begin{scope}[shift={(0,0)}]
  \draw[->] (0,0) -- (6.2,0) node[below] {$x$};
  \draw[->] (0,-2.2) -- (0,2.2) node[left] {$U(x)$};

  \draw[thick]
    plot[domain=0:6,samples=150,smooth]
      (\x,{0.7*sin(5*\x r)+0.4*sin(11*\x r)+0.25*sin(19*\x r)});

  \node at (3,-2.0) {Uncorrelated / highly rugged};
\end{scope}

\begin{scope}[shift={(8,0)}]
  \draw[->] (0,0) -- (6.2,0) node[below] {$x$};
  \draw[->] (0,-2.2) -- (0,2.2) node[left] {$U_{\rm eff}(x)$};

  \draw[thick]
    plot[domain=0:6,samples=150,smooth]
      (\x,{0.9*sin(3*\x r)+0.4*sin(5*\x r)});

  \fill[gray!20,rounded corners] (2.0,-2.1) rectangle (3.2,2.1);
  \draw[gray!70,thick,rounded corners] (2.0,-2.1) rectangle (3.2,2.1);
  \node[gray!50!black] at (2.6,1.7) {local};
  \node[gray!50!black] at (2.6,1.4) {averaging};

  \node at (3,-2.0) {Correlated / locally smoothed};
\end{scope}

\end{tikzpicture}
\caption{Schematic illustration of the effect of spatial correlations and local
smoothing on a rugged potential.  
Left: an ``uncorrelated'' rugged landscape with strong, rapidly varying
fluctuations in $U(x)$.  
Right: a correlated (or locally averaged) landscape in which high-frequency
roughness has been smoothed out over a small window, as assumed in Zwanzig's
local-averaging (cumulant-expansion) treatment.  
The figure is schematic and not based on any specific numerical data.}
\label{fig:smoothing_schematic}
\end{figure}
\subsection {Quantitative approach to Zwanzig in the BSB simulations}

Banerjee, Seki, and Bagchi carried out Brownian dynamics simulations of a
particle diffusing on (i) an uncorrelated Gaussian random lattice and
(ii) a continuous Gaussian field with built–in spatial correlations.\cite{banerjee_seki_biswas_bagchi_spatial}
For the uncorrelated lattice, they found that Zwanzig’s expression
for the diffusion coefficient,
\[
D_{\mathrm{Zw}} = D_0 \exp(-\beta^2 \epsilon^2),
\]
\emph{overestimates} the simulated diffusion coefficient by nearly an order
of magnitude at moderately high ruggedness (for example, when
$\epsilon/k_B T$ is of order $3$–$4$).\cite{banerjee_seki_biswas_bagchi_spatial}
In other words, the ratio
\[
\frac{D_{\mathrm{sim}}}{D_{\mathrm{Zw}}}
\]
is of order $10^{-1}$ in the uncorrelated case, consistent with the strong
slowing down caused by three-site traps.

In striking contrast, when the same authors considered a \emph{continuous}
Gaussian random field with finite spatial correlation length (constructed
by superimposing Gaussian modes in real space), the simulated diffusion
coefficients were found to lie essentially on Zwanzig’s original
prediction over the same range of $\epsilon$.\cite{banerjee_seki_bagchi_jcp}
Within numerical uncertainty, one obtains
\begin{equation}
\frac{D_{\mathrm{sim}}}{D_{\mathrm{Zw}}} \approx 1
\end{equation}
for the correlated continuous field, demonstrating that spatial
correlations smooth out the pathological three-site traps and restore
the validity of the mean-field, cumulant-expansion result.
These simulation data provide a concrete quantitative confirmation of the
mechanism discussed in this work: uncorrelated Gaussian disorder produces
anomalously small diffusion, while Gaussian spatial correlations
eliminate rare deep traps and recover Zwanzig’s exponential scaling.
\begin{table}[t]
\centering
\caption{Comparison of uncorrelated and spatially correlated rugged landscapes.
Energies are in units of $k_B T$.  ``TST'' denotes a three-site trap 
$E_{i-1}, E_i, E_{i+1}$ with a deep central well.}
\vspace{6pt}
\begin{tabular}{|c|c|c|}
\hline
\textbf{Property} 
& \textbf{Uncorrelated Energies} 
& \textbf{Spatially Correlated Energies} \\
\hline
Typical TST triplet  
& $(4,\,-3,\,5)$  
& $(1.2,\,-2.5,\,0.9)$ \\
\hline
Escape barriers  
& $7$ and $8$  
& $3.7$ and $3.4$ \\
\hline
Escape rate $k_{\rm esc}$  
& $1.25\times 10^{-3}\,k_0$  
& $5.83\times 10^{-2}\,k_0$ \\
\hline
Escape time $\tau_{\rm esc}$  
& $\approx 8.0 \times 10^{2}\,k_0^{-1}$ 
& $\approx 1.7 \times 10^{1}\,k_0^{-1}$ \\
\hline
Severity of TSTs  
& Very high: deep asymmetric traps 
& Mild: asymmetric traps suppressed \\
\hline
BSB simulation result  
& $D_{\rm sim}/D_{\rm Zw} \sim 10^{-1}$  
& $D_{\rm sim}/D_{\rm Zw} \approx 1$ \\
\hline
Dominant physics  
& Rare deep traps control transport  
& Landscape is smooth for mean-field theory \\
\hline
Outcome  
& Zwanzig \emph{fails}  
& Zwanzig is \emph{restored} \\
\hline
\end{tabular}
\label{table:comparison}
\end{table}

\section{Implications and Outlook}

The central outcome of this work is that spatial correlations fundamentally alter transport in disordered landscapes.  
Uncorrelated roughness produces pathological trapping dominated by rare events, but Gaussian correlations suppress extreme asymmetry and restore Zwanzig’s simple exponential scaling.

In biological systems such as enzymes diffusing along DNA,\cite{blainey_xie_nsmb} correlated roughness is physically natural: the sequence-dependent binding energy varies smoothly over several base pairs.  
Similarly, in polymers, glasses, and proteins, extended interactions generate intrinsic correlations in the energy landscape.
Gaussian spatial correlations, originally developed in astrophysics to describe
turbulent density fluctuations, provide a natural mechanism for smoothing a rugged potential.  
They reduce the variance of roughness increments, correlate successive increments,
suppress the probability of deep asymmetric multi-site traps, and restore the
validity of Zwanzig’s mean-field exponential scaling.  
The central insight is that \emph{correlated} roughness behaves smoothly enough for
coarse-grained mobility averaging to hold, whereas \emph{uncorrelated} roughness is
dominated by rare but "catastrophic traps" that invalidate any mean-field approach.
%

Thus, the BSB extension shows that rugged landscape diffusion is governed not only by the magnitude of disorder but by its \emph{spatial structure}.  
Gaussian correlations---first introduced in astrophysics
\cite{gaussian_astrophysics1, gaussian_astrophysics2}---provide an elegant and quantitative framework for understanding this effect.


\bibliographystyle{unsrt}

\end{document}